%% file: main.tex
\documentclass[conference]{IEEEtran}
\IEEEoverridecommandlockouts
\usepackage{cite}
\usepackage{amsmath,amssymb,amsfonts}
\usepackage{algorithmic}
\usepackage{graphicx}
\usepackage{textcomp}
\usepackage{xcolor}
\def\BibTeX{{\rm B\kern-.05em{\sc i\kern-.025em b}\kern-.08em
    T\kern-.1667em\lower.7ex\hbox{E}\kern-.125emX}}

\makeatletter
 \let\old@ps@headings\ps@headings
 \let\old@ps@IEEEtitlepagestyle\ps@IEEEtitlepagestyle
 \def\confheader#1{%
 \def\ps@IEEEtitlepagestyle{%
 \old@ps@IEEEtitlepagestyle%
 \def\@oddhead{\strut\hfill#1\hfill\strut}%
 \def\@evenhead{\strut\hfill#1\hfill\strut}%
 }%
 \ps@headings%
 }
\makeatother

\confheader{%
\parbox{18cm}{\centering Accepted copy for Publication at the IEEE  International Conference on RFID-Technologies and Applications, 2021 \\ 
Final published version available at: https://doi.org/10.1109/RFID-TA53372.2021.9617320}}
    
\begin{document}

\title{Towards Trustworthy NFC-based Sensor Readout for Battery Packs in Battery Management Systems
}

\author{
\IEEEauthorblockN{Fikret Basic, Martin Gaertner, Christian Steger}
\IEEEauthorblockA{\textit{Institute for Technical Informatics} \\
\textit{Graz University of Technology}\\
Graz, Austria \\
\{basic, gaertner, steger\}@tugraz.at}
}

\maketitle

\begin{abstract}
In the last several years, wireless Battery Management Systems (BMS) have slowly become a topic of interest from both academia and industry. It came from a necessity derived from the increased production and use in different systems, including the electrical vehicles. Wireless communication allows for a more flexible and cost-efficient sensor installation in battery packs. However, many wireless technologies, such as those that use the 2.4 GHz frequency band, suffer from the interference limitations that need to be addressed. In this paper we present an alternative approach to communication in BMS that relies on the use of Near Field Communication (NFC) technology for battery sensor readouts. Due to a vital concern over the counterfeited battery pack products, security measures are also considered. To this end, we propose the use of an effective, and easy to integrate, authentication schema that is supported by the dedicated NFC devices. To test the usability of our design, a demonstrator using the targeted devices was implemented and evaluated.
\end{abstract}


\section{Introduction}
\input{introduction}

\section{Background and Related Work}
\input{related_work}

\section{Design of the Novel BMS NFC Sensor Readout}
\input{design}

\section{Evaluation}
\input{evaluation}

\section{Conclusion and Future Work}
\input{conclusion}

\section*{Acknowledgment}
This project has received funding from the ``EFREtop: Securely Applied Machine Learning - Battery Management Systems'' (Acronym ``SEAMAL BMS'', FFG Nr. 880564).

\bibliographystyle{ieeetr}
\bibliography{references}

\end{document}

%% file: introduction.tex
Over the years, Battery Management Systems (BMS) have seen an increased interest in the research community, primarily due to the higher digitization and use in different applications. They play an important role in many systems, but are often mentioned today as battery control devices used in smart power grids, and electric or hybrid vehicles~\cite{8168251}. BMS are mainly used to handle the balancing of the large battery cells during charging \& discharging cycles, as well as to offer diagnostic services to track their lifetime usage~\cite{andrea2010battery}.

A BMS can be deployed in different topologies and usually consists of various devices. They generally contain a main BMS controller, which in a modulated setting, communicates with individual Battery Cell Controllers (BCC). The BCCs help in relaying diagnostic data back to the BMS controller through monitoring and control of individual battery cells. Data received from these sensors is critical in preventing dangerous incidents like the thermal runaway, which happens due to the rapid increase in battery temperature~\cite{fireBattery}. However, a BMS controller can only act as long as it has the correct information on the current state inside the battery pack, i.e. it is dependant on the sensor readouts. Two main factors influence the accuracy of these readings: (i) the number of the sensors used, and (ii) the relative position of the sensors to their target of interest. 

Commonly, BMS use wired connection to handle the communication with battery cell sensors. This imposes three main limitations however: 
\begin{enumerate}
    \item \textit{Assembly cost:} Each connection to and from the cell and sensor source needs to be physically soldered, and it needs to account for materials used.
    \item \textit{Scalability:} The complexity of design, deployment, and afterwards maintenance, boosts with the increase of the number of battery cell sensors. 
    \item \textit{Area coverage:} Due to the physical wires used, sensor placement is often limited. Often, this makes the use of certain areas impossible, e.g. inner housing of the cells.
\end{enumerate}

In order to alleviate the aforementioned wired limitations, it is possible to replace wired with wireless technology networks. However, we see several challenges that need to addressed when choosing an appropriate wireless technology:
\begin{itemize}
    \item \textit{Restricted data throughput}: For a safe and continuous execution of the BMS operations, it is necessary to maintain a steady and fast flow of data from the battery cell sensors to the BCCs and afterwards the BMS controller.
    \item \textit{Interference}: The use of the same frequency band, even through different communication technologies, will cause interference. This is especially true with the 2.4 GHz band which is used by several technologies, most prominent being LR-WPAN, Bluetooth, ZigBee, WiFi.
    \item \textit{Multipath propagation}: Maintaining communication sight and reliability under strict and obstructive environments.
    \item \textit{Security concerns:} Unless placed in an enclosed case, wireless networks are prone to eavesdropping, remote attacks, and other malicious incursions~\cite{9090905}.
\end{itemize}

As an answer to the mentioned obstacles, we propose the realization of the communication between the battery cell sensors and the BCCs to be done using NFC. By employing NFC, we are not only able to answer to the design restrictions imposed through the use of the wired communication, but also address the challenges introduced when using a wireless communication. Furthermore, we address a general safety BMS requirement centered around the battery cells source validity~\cite{8813669, 101093}. It is important that only battery cells that come from valid and approved manufactures are installed, as inadequate battery cells can potentially result in hazards that cannot be mitigated by the BMS controller.\\
\textbf{Contributions.} Summarized, our main contributions contained in this paper are: (i) We present an NFC-based approach that can be used for BMS sensor readout, specifically the communication between the battery cell sensors and the battery cell controllers. (ii) To counter potential malicious and counterfeiting attempts, we provide a security solution that can be easily integrated. (iii) The implementation of the proposed design and evaluation of its utility using a BMS test system.

%% file: related_work.tex
With the increase of the number of battery cells in modern BMS, new topologies and architectures had to be introduced. A focus was set in using derivations of modular and distributed BMS~\cite{andrea2010battery}. 
This all further lead to an increase in expenses and complexity in cable installation. Different models have been proposed based on the wireless technology used, such as the use of Bluetooth~\cite{7151581, 7357002}, and ZigBee~\cite{Rahman_2017}. They primarily focus on the communication between the BCCs and the main BMS controller using these wireless technologies. We extend the wireless usage by focusing on the BCC and sensor communication through NFC utilization.

The use of the NFC technology in larger system infrastructures has already been investigated before. Specifically, research presented by Ulz et al. \cite{8098906} proposes the use of the NFC-based communication for robot-machine interaction in a Industry 4.0 setting. Additionally, work by Chen et al.~\cite{6228335} investigates secure authentication and anti-counterfeiting methods using RFID. Alzahrani et al.~\cite{9075232} proposes an NFC-focused anti-counterfeiting system. Despite the large amount of research being done both for the general wireless BMS and the integration of NFC in similar environments, not many specific work has yet been done that combines these two fields of interest. Work done by Schneider et al.~\cite{6229439} focuses largely in this field, by also proposing a design approach for wireless BMS battery sensors utilizing the same RFID technology. However, one of the main focal points in that paper is placed on the issues caused through the galvanic isolation. Moreover, due to the date when the paper was published, it does not account for the newer BMS modular architectures and modern NFC derivations, alongside the security aspects. In this work, we try to bridge that gap and show the potential of using NFC in a hard-to-reach sensor environments, while at the same time giving attention to the security requirements. 

%% file: design.tex
\begin{figure}[htbp]
  \centering
  \includegraphics[width=0.85\linewidth]{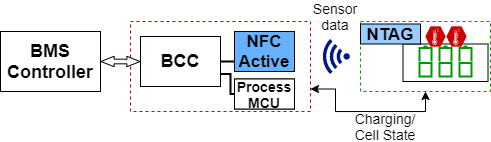}
  \caption{Proposed BMS modular design architecture utilizing NFC components.}
  \label{fig:bms_design_architecture}
\end{figure}

For our targeted design architecture we divide the entire system into three main modules: (i) a \textit{BMS controller}, (ii) a \textit{Cell control board}, (iii) and a \textit{Battery module}. The BMS controller can either contain one or multiple different \textit{Microcontroller Units} (MCUs) set for the overall BMS control and status monitoring. It communicates with the cell control board that contains a BCC, the NFC reader as the communication interface, and optionally an additional control MCU for the protocol handling. The battery module contains battery cells, sensors, and NFC communication interface to the cell board. In our case, this interface is a \textit{NFC-Tag} (NTAG). This is illustrated in Figure~\ref{fig:bms_design_architecture}. For charging \& discharging cycles, as well as the related voltage readings, we still rely on the hardwired measurements from conventional BMS designs.

\subsection{NFC Communication}
For the BCCs and battery cells to be able to communicate using the wireless NFC technology, appropriate devices and communication mode need to be chosen. In the presented design this is done over the \textit{Reader/Writer mode}. The NFC reader is the active device which is connected to a specialized controller BCC, as well as to an MCU for pre-processing and security operations. Before the communication begins, the NFC reader needs to have discovered the necessary NTAG(s) using a discovery loop. Afterwards, the authentication process starts. Following it, the NTAG proceeds to initiate self-configuration and prepares to communicate with both the sensor and the NFC reader. Since in a standard environment the same devices are going to be used also for the subsequent measurement readings, the initialization and configuration steps can be cached and therefore omitted. 

\subsection{Energy Harvesting and Positioning}
A disadvantage that NFC has over most other wireless technologies is in its relatively short range. In the presented design this is of no issue, as the BCCs and battery cells are usually tightly packed and installed together.
The NFC in our design uses the energy harvesting feature to power up the NTAG from the reader. This feature limits the distance between the antennas. Depending on the environment, the distance peaks approximately at $5.4\,cm$. For a feasible communication, and optimal initialization time, we opted to use a distance of $2\,cm$.
As both the sensor and the NTAG reside on the battery module, it would be possible for them to be directly powered as it is done in a conventional design. However, this characteristic is not present in our design, as using the wiring to the battery modules would violate one of the design requirements set on reducing the extent of the necessary wires.

\subsection{Authentication Protocol}
In terms of the security, an advantage that the NFC has over the use of other wireless technologies is in both its short range and frequency band. This limits the list of technologies that a potential attacker could use to attack the system. Since battery modules are usually enclosed in a protective case together with the BCC, this means that the main potential attack vector on these modules would be one through \textit{counterfeiting}.

\begin{figure}[htbp]
  \centering
  \includegraphics[width=0.50\linewidth]{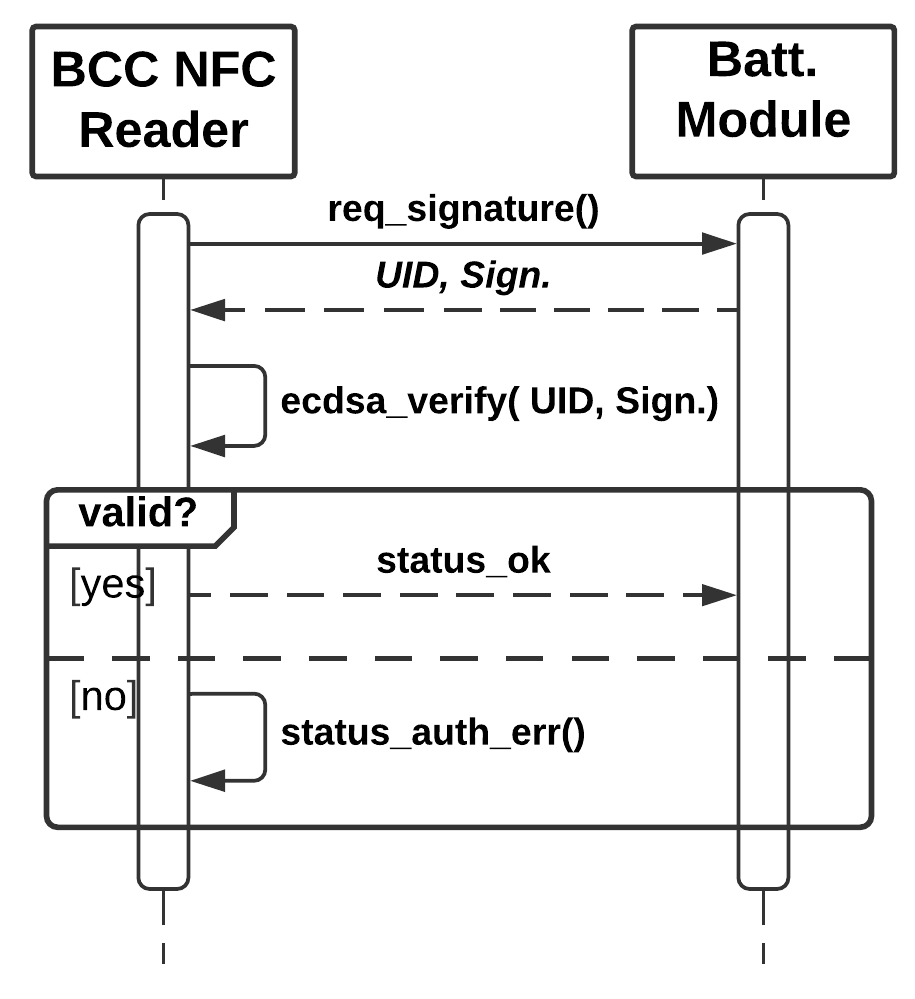}
  \caption{Sequence diagram of the authentication protocol.}
  \label{fig:bms_nfc_sequence_diag}
\end{figure}

To be able to securely verify that the battery modules are valid, we integrate the use of an authentication protocol in our design. This is achieved by verifying a value that needs to be unique to each device. Since NTAGs are usually shipped with an Unique Identifier (UID) value, we can use it as an input for an Elliptic Curve Digital Signature Algorithm (ECDSA). In our design, we use the \textit{secp128r1} protocol as the Elliptic Curve (EC) function, having a good balance between the performance and output sizes. The signature value, which is calculated with a private key during the manufacturing process or subsequently updated, is then stored in a protected memory space located on the NTAG chip. The BCC needs to have access to the public key, either it being pre-embedded, or accessed through other secure channels. The authentication protocol is shown in Figure~\ref{fig:bms_nfc_sequence_diag}. Before the signature verification takes place, the UID validity is first checked against the list of valid devices.

%% file: evaluation.tex
\subsection{Test System Implementation}
For the purpose of testing our design approach, we implemented a test suite which contains the necessary BMS modules, as well as the additional NFC equipment. We aimed to use the NFC modules which support the latest \textit{NFC Type 5 Tag} technology. Furthermore, the used components are automotive-graded where applicable, for the purpose of replicating a real-world use-case as closely as possible. To that end, all devices used, except of the temperature sensor, come from the NXP Semiconductors lineup of products. 

\begin{figure}[htbp]
  \centering
  \includegraphics[width=0.95\linewidth]{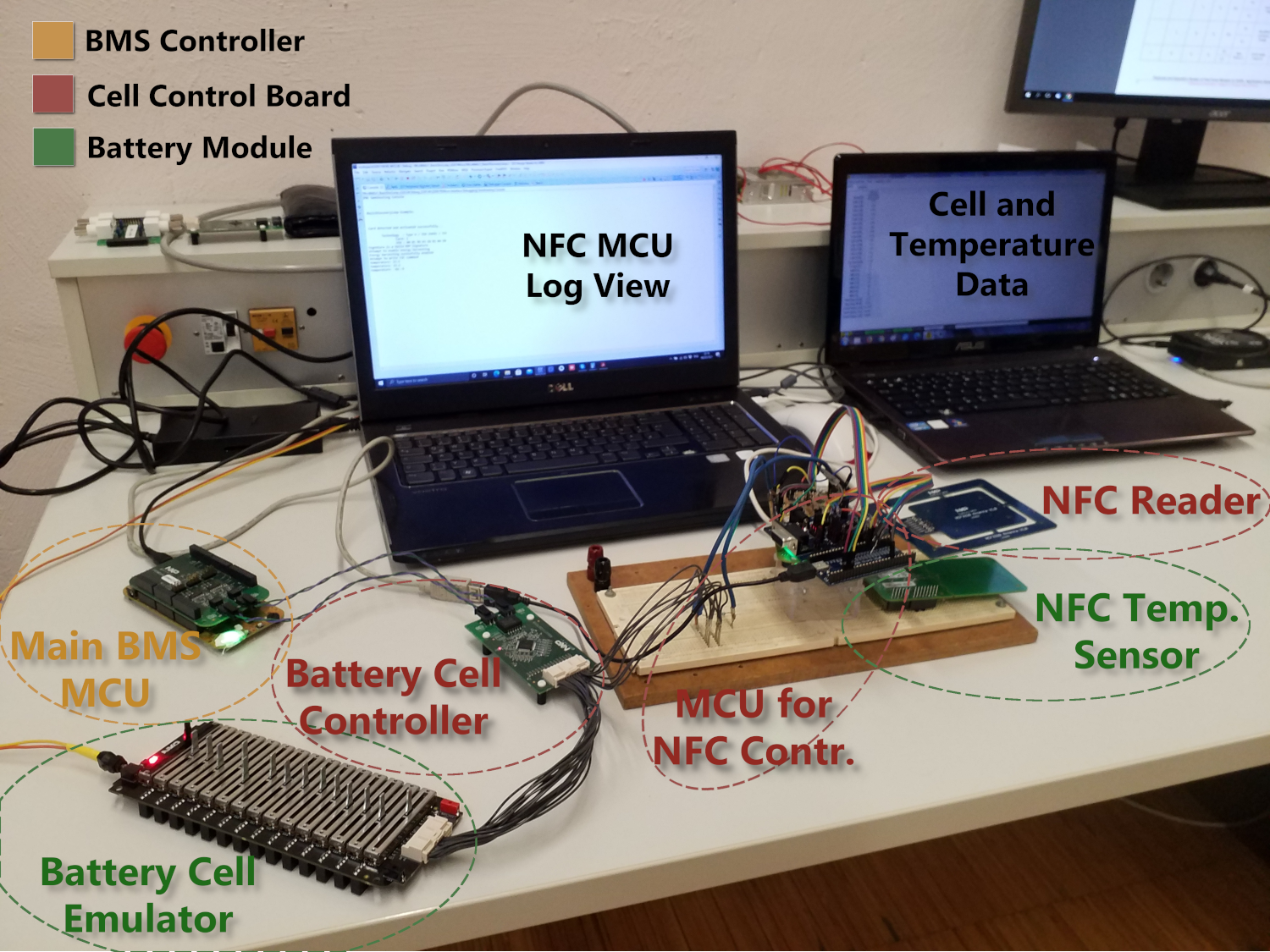}
  \caption{Test setup for the BMS NFC sensor readout.}
  \label{fig:photo_test_setup}
\end{figure}

As the main BMS controller we use a S32K144. It communicates with the cell control board via the FRDMDUAL33664 shield. It is further connected to a RD33771CDST that houses a MC33771C which functions as a BCC. The cell control board contains an automotive NFC Reader for handling the NFC transmissions, and an another S32K144 as the MCU for the purpose of programming and testing. The battery module consists of a BATT-14CEMULATOR that serves as a battery emulator, an NTAG component as the passive NFC device, and a BMP180 temperature sensor. The BCC is able to receive the emulated cell voltage data from the battery emulator, while the temperature sensor data is sent through the NFC interface. Both the temperature and the cell voltage data are first received by the BCC and then transmitted to the BMS controller. For the authentication protocol, we base our implementation on the \textit{originality signature} feature found on the NXP's RFID devices. Signature calculation and verification are handled via the \textit{ecc-nano} library. Elements of the development and evaluation were handled in a recent master's thesis~\cite{mastersthesis}. The system is shown in Figure~\ref{fig:photo_test_setup}.

\subsection{Time Measurements}
\label{sec:time_meas}
\begin{figure}[htbp]
  \centering
  \includegraphics[width=0.90\linewidth]{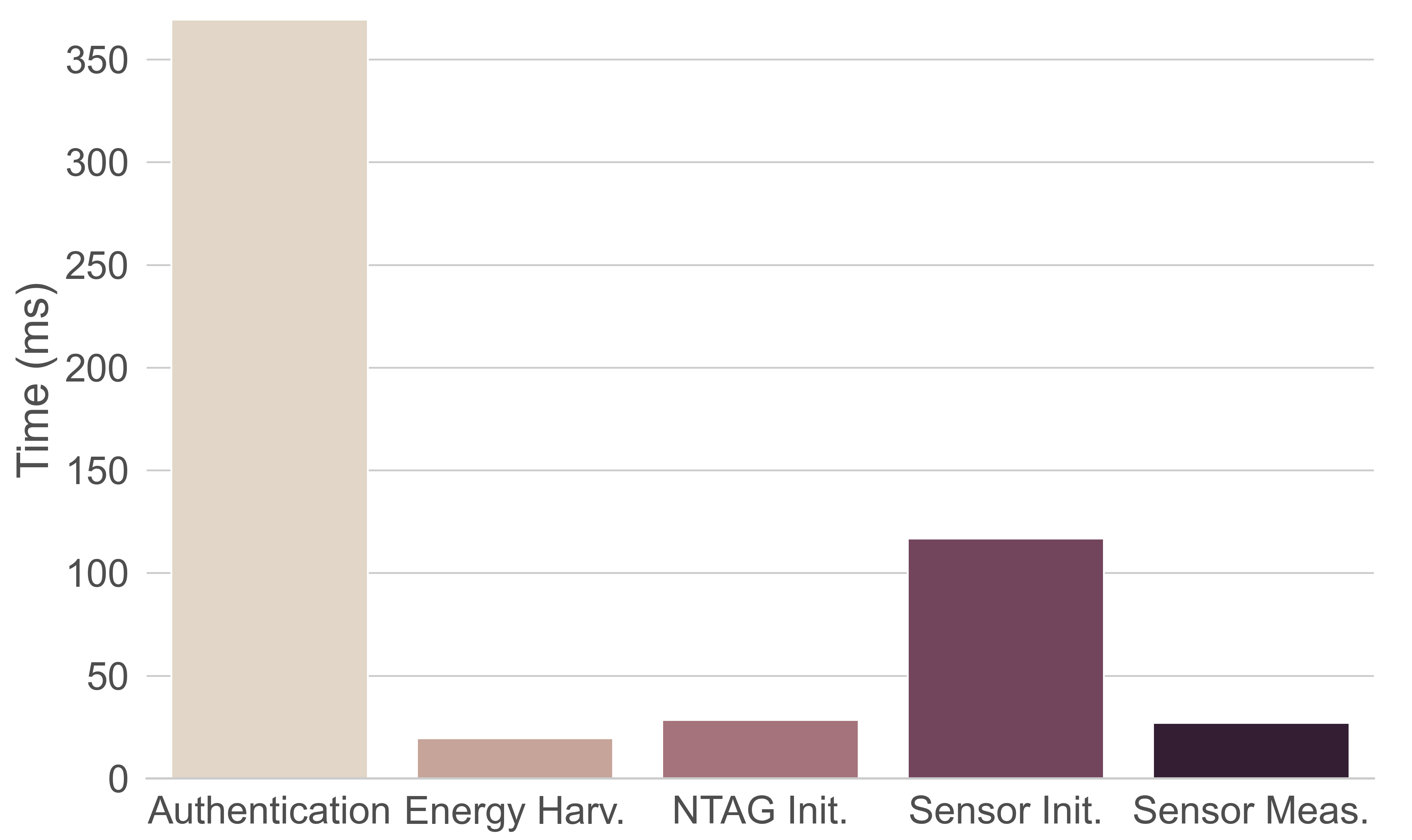}
  \caption{Time measurement results for (a) Init. phase (Authentication, Energy Harv., NTAG Init., Sensor Init.) and (b) Monitoring phase (Sensor Meas.).}
  \label{fig:bms_time_measurements}
\end{figure}

We divided the entire BMS monitoring process into two phases: (i) \textit{Initialization phase:} executed only once for device preparation and configuration, and (ii) \textit{Monitoring phase:} continuous action that is called on every sample step to measure and retrieve sensor and cell data. Individual steps, as well as their time measurements, are shown in Figure~\ref{fig:bms_time_measurements}. 

We start the process after the NTAGs have already been discovered. As the first step the \textit{Authentication protocol} is run. This protocol run includes both sending an authentication request from the NFC reader, the response from the NTAG, and the verification calculation on the MCU that is connected with the NFC reader. The authentication step showed an average time of $369.3\,ms$, with majority of it being spent on the verification process. With the NTAG verified, the \textit{Energy harvesting} check is handled which lasts for $19.64\,ms$. Finally, the NTAG operation initialization is run which measured $29.16\,ms$, followed with the sensor initialization that took $116.1\,ms$. 

After the initialization phase is finished, there is no need to reconfigure the devices during the system run. For the monitoring phase, \textit{Sensor Measurements} are read and transmitted to the BCC using NFC communication. This phase is repeatable, with each action showing a time of $27.2\,ms$. 

\subsection{Security Threat Analysis}
To evaluate the achieved security protection, a threat analysis was conducted. We demonstrate a summarized representation of the analysis based on the \textit{Threats} (T), \textit{Assets} (A), and \textit{Countermeasures} (C). In our security model, we argue the following assumptions: (i) A battery module can only be communicated with via an adequate BCC, (ii) Both the cell control board and the battery module are enclosed in a chassis and the external communication can only be achieved through the BMS controller, (iii) Every newly added and unknown battery module is considered untrustworthy.

\begin{figure}[htbp]
  \centering
  \includegraphics[width=0.92\linewidth]{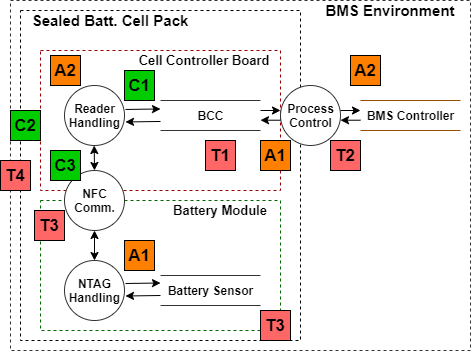}
  \caption{Data Flow Diagram representation of the security analysis.}
  \label{fig:thread_model}
\end{figure}

To conduct our security analysis, we have used \textit{Data Flow Diagram} (DFD) to illustrate the system threat model as seen in Figure~\ref{fig:thread_model}. Here, we indicate all assets, threats and countermeasures, as well as points of their potential impact. In our use-case we have two assets that we want to protect: (A1)~\textit{Sensor (status) data}: data retrieved from the cell sensors and handled throughout the system, and (A2)~\textit{System integrity}: both hardware and software integrity and correctness. To disturb the cell balancing control, we indicate (T1)~\textit{Battery control obstruction} as a potential threat. Further, an attacker might, through data modification, try to (T2)~\textit{Tamper with BMS status messages}. Both of these threats can be protected against with the implemented (C1)~\textit{Authentication through signature validation} countermeasure. Another possibility would be for an attacker to gain a (T3)~\textit{Backdoor access} through either the NFC interface or the counterfeited battery module. This is again protected by (C1), but also for the NFC by (C3)~\textit{NFC Physical layer characteristic} for which the attacker could not mount an attack of such proximity. Lastly, a (T4)~\textit{Remote attack} could also be launched from the outside using a wireless communication. In this case, we indicate that both the (C2) \textit{Cell pack sealing} and also (C3) would hamper the possibility of such an attack.

%% file: conclusion.tex
In this paper we present an idea of using NFC as a wireless communication interface for battery sensor readouts in BMS. To alleviate the risk of the counterfeited battery cells and prevent safety and security threats, an authentication model was proposed and evaluated. Experimental results using real components showed the feasibility of our approach, but also challenges related to the timing optimization, and antenna or sensor placement. For the future work, we plan to evaluate the design using different antenna orientations, and also consider a security protocol extension for the rest of the BMS. 